\documentclass{article}
\usepackage{color}
\usepackage[english]{babel}
\usepackage{psfrag}
\usepackage{epsfig}
\usepackage{latexsym}
\usepackage{graphics}
\usepackage{verbatim}
\usepackage{mathrsfs}
\usepackage{lscape}
\begin{document}
\newcommand{\be}{\begin{equation}}
\newcommand{\ee}{\end{equation}}
\newcommand{\bestar}{\[}
\newcommand{\eestar}{\]}
\newcommand{\beastar}{\begin{eqnarray*}}
\newcommand{\eeastar}{\end{eqnarray*}}
\newcommand{\beq}{\begin{equation}}
\newcommand{\eeq}{\end{equation}}
\newcommand{\bea}{\begin{eqnarray}}
\newcommand{\eea}{\end{eqnarray}}
\newcommand{\disp}{\displaystyle}
\newcommand{\mbf}{\mathbf}
\newcommand{\dint}{\displaystyle\int}
\renewcommand{\u}{\underline}
\renewcommand{\o}{\overline}
\newcommand{\bi}{\begin{itemize}}
\newcommand{\ei}{\end{itemize}}
\newcommand{\bfig}{\begin{figure}[htb]\begin{center}}
\newcommand{\efig}{\end{center}\end{figure}}
\newcommand{\eqref}[1]{(\ref{#1})}
\newcommand{\eqm}[2]{~(\ref{#1}-\ref{#2})}
\newcommand{\eq}[1]{~(\ref{#1})}
\newcommand{\eqq}[2]{~(\ref{#1},\ref{#2})}
\newcommand{\eqqq}[3]{~(\ref{#1},\ref{#2},\ref{#3})}
\newcommand{\order}{{{\mathcal O}}}
\newcommand{\text}{\mbox}
\newcommand{\ie}{{\it i.e.}}
\newcommand{\eg}{{\it e.g.}}
\newcommand{\rd}{\color{red}}
\newcommand{\bl}{\color{blue}}
\newcommand{\gr}{\color[rgb]{0.333333,0.752941,0.203922}}
%
%
%
%
\newcommand{\kboltz}{\kappa_{\rm B}}
\newcommand{\ext}{^{\rm ext}}
\newcommand{\f}{^{\rm f}}
\newcommand{\G}{^{\rm G}}
\renewcommand{\L}{^{\rm L}}
\newcommand{\g}{^{\rm g}}
\renewcommand{\l}{^{\rm l}}
\newcommand{\W}{^{\rm W}}
\newcommand{\ph}{^a}
\newcommand{\parent}{^{(0)}}
\newcommand{\tot}{_{\rm{tot}}}
\newcommand{\src}{\psi}
\newcommand{\out}{^{\rm out}}
\newcommand{\vap}{^{\rm vap}}
\newcommand{\cond}{^{\rm cond}}
\newcommand{\esp}{_{\rm exp}}
\newcommand{\LG}{_{\rm LG}}
\newcommand{\WG}{^{\rm WG}}
\newcommand{\WL}{^{\rm WL}}
\newcommand{\Mol}{N}
\newcommand{\mol}{n}
\newcommand{\vectrho}{{\text{\boldmath$\rho$}}}
\newcommand{\vectMol}{{\text{\boldmath$\Mol$}}}
\newcommand{\vectmol}{{\text{\boldmath$\mol$}}}
\newcommand{\molfrac}{x}
\newcommand{\vol}{v}
\newcommand{\Vol}{V}
\newcommand{\vel}{w}
\newcommand{\surf}{A}
\newcommand{\Ent}{H}
\newcommand{\fen}{f}
\newcommand{\Fen}{F}
\newcommand{\Fid}{F_{\rm id}}
\newcommand{\fid}{f_{\rm id}}
\newcommand{\fmom}{f_{\rm mom}}
\newcommand{\fexc}{\tilde{f}}
\newcommand{\Fexc}{\widetilde F}
\newcommand{\Gibbs}{G}
\newcommand{\gibbs}{g}
\newcommand{\gid}{\gibbs_{\rm id}}
\newcommand{\gexc}{\tilde{\gibbs}}
\newcommand{\entr}{s}
\newcommand{\Entr}{S}
\newcommand{\Inte}{U}
\newcommand{\ent}{h}
\newcommand{\inte}{u}
\newcommand{\temp}{T}
\newcommand{\pres}{P}
\newcommand{\heat}{Q}
\newcommand{\heatf}{q}
\newcommand{\comp}{z}
\newcommand{\calv}{c_\vol}
\newcommand{\calp}{c_\pres}
\newcommand{\crit}{_{\rm c}}
\newcommand{\hcoeff}{\lambda}
\newcommand{\noz}{_{\rm noz}}
\newcommand{\visco}{\eta}
\newcommand{\tcond}{\kappa}
\title{Phase equilibria of polydisperse hydrocarbons: moment free energy 
  method analysis}
\author{A. Speranza$^{(*)}$, F. Dipatti$^{(**)}$ and A. Terenzi$^{(***)}$\\
  {\em $(*) (**)$ Industrial Innovation Through Technological 
    Transfer, I$^2$T$^3$ Onlus}\\
    {\em Dip. di Matematica Universit\`a di Firenze}\\
    {\em V.le Morgagni 67/a Firenze, Italia}\\
    {\em E-mail} {\tt alessandro.speranza@i2t3.unifi.it}\\
    {\em $(***)$ Snamprogetti S.p.A. Via Toniolo 1, Fano (PU), Italia}}
\date{}
\maketitle
\begin{abstract}
  We  analyze   the  phase  equilibria  of   systems  of  polydisperse
  hydrocarbons  by means  of  the recently  introduced moment  method.
  Hydrocarbons   are  modelled   with   the  Soave-Redlick-Kwong   and
  Peng-Robinson  equations  of   states.  Numerical  results  show  no
  particular  qualitative  difference  between  the two  equations  of
  states. Furthermore,  in general the  moment method proves to  be an
  excellent  method  for  solving  phase  equilibria  of  polydisperse
  systems,  showing  excellent  agreement  with  previous  results  and
  allowing a  great improvement in generality of  the numerical scheme
  and speed of computation.
\end{abstract}
\section{Introduction}
In this paper we analyze the phase behaviour of a mixture of
hydrocarbons, by means of the moment
method~\cite{SolCat98,SolWarCat01}. This method allows to reduce the
number of degrees of freedom of the free energy, which normally
depends on the concentration of each specie in the mixture, to a
smaller number of moments of the density distribution which already
appear in the excess part of the free energy. By doing this, one is
able is reduce the number of equations needed to
analyze the phase equilibria and, at the same time, by projecting the
free energy onto the space generated by the moments only, to check for
global and local stability of the phases~\cite{SolWarCat01}. 

The approximation made when introducing the moment free energy can be
efficiently controlled and minimized by means of the adaptive method
of choice of extra moments~\cite{SpeSol_p2}, which allows to reduce
the deviation of the moment method solution from the exact solution, by
simply retaining two extra moments, beyond the ones appearing in the
excess free energy. This iterative method, which it can be proven,
converges to the exact solution, as long as it converges at all, shows
to represent an excellent compromise between approximation, which can
easily be reduced to an error smaller than 0.01\%, and computational
speed. Furthermore, the resulting algorithm turns out to be stable and
to be very little affected by the number of species in the mixture. In
fact, as the number of unknowns is not increased by the increase of
the number of species, the computation is hardly affected at all, with
just a small influence on its global speed of computation, while no
relevant error is introduced. 

The numerical results agree very well with the results obtained with a
widely diffused commercial program in different points of the phase
diagram. The concentration of each component in the coexisting phases
and the density of both phases are evaluated correctly. Clud point is
detected exactly. Furthermore, the introduction of heavy species, up
to n-C15, even present in very small amount, does not compromise either numerical
results or computation. 

\section{Polydisperse hydrocarbons}
In order to analyze the phase equilibria of hydrocarbons, we will refer
to the two equations of state most widely used to describe them, \ie,
the Soave-Redlick-Kwong (SRK)~\cite{Soave72} equation of state and the Peng-Robinson
(PR)~\cite{PenRob76} equation of state. Both are cubic equations of
state and thus are able to predict gas-liquid phase
transitions. Although originally introduced for pure systems, as we
will see, they are both easily extended to describe multicomponent,\ie,
polydisperse systems. As we will show, given the polydisperse form of two equations of
state, one can easily obtain the Gibbs and Helmoltz free energies, by
Legendre transforming, and therefore obtain the phase equilibrium
equations that are to be solved, in order to fully analyze the phase
behaviour of the system.

The SRK equation of state is generally written as
\beq\label{eq:srk}
\pres=\frac{\Mol\kboltz\temp}{\Vol-b}-\frac{\Mol^2\alpha(\temp) a_{C}}{\Vol\left(\Vol+b\right)}
\eeq
where $\Mol$ is the total number of particles, $\Vol$ the total
volume, $\kboltz$ the Boltzman constant and the parameters $a,\ b$ and $\alpha(\temp)$
depend on the critical temperature and pressure, shape and 
size of the molucules etc., of the specific hydrocarbon.

The extension of the equation above to the case of polydisperse system
is rather straightforward, if one introduces a set of parameters
$a_{C,i},\ b_i,\ \alpha_i(\temp)$ for each specie $i$ and defines new
global parameters $B$ and $D$ as follows 
\beq\label{eq:B_poly}
B=\sum_i\Mol_i b_i
\eeq
and
\beq\label{eq:D_poly}
D(\temp)=\sum_{i,j}\Mol_i\Mol_ja_{ij}(\temp)
\eeq
where 
\beq\label{eq:a_poly}
a_{ij}(\temp)=\sqrt{a_{C,i}a_{C,j}\alpha_i(\temp)\alpha_j(\temp)}
\eeq
In this way, the polydisperse version of Eq.~\eqref{eq:srk} is
\beq\label{eq:srk_poly_gen}
\pres=\frac{\Mol \kboltz \temp}{\Vol-B}-\frac{D}{\Vol(\Vol+B)}
\eeq
where $\Mol=\sum_i\Mol_i$ is the total number of particles of the
system.

The Helmoltz free energy can now be obtained simply by solving the
equation
\[
\pres=-\frac{\partial\Fen}{\partial\Vol}
\]
and by introducing $\Fid$, the ideal part of the free energy, \ie, the
free energy of a mixture of ideal gas
\beq\label{eq:free_id}
\Fid=\kboltz\temp\sum_i\Mol_i\left(\ln\frac{\Mol_i}{\Vol}-1\right)
\eeq
The free energy then turns out to be
\beq\label{eq:free_poly}
\Fen({\mbf
  n},\Vol,\temp)=\Fid+\Mol\kboltz\temp\ln\frac{\Vol}{\Vol-B}-\frac{D}{B}\ln\frac{V+B}{V}
\eeq
The above quantity is extensive, one can therefore define an
intensive ``free energy density'' $\fen=\Fen/Vol$. Introducing a density
distribution $\rho(k)=\Mol_k/\Vol$ and multiplying by
$\beta=1/\kboltz\temp$, the free energy density turns out to be
\beq\label{eq:free_srk}
\beta\fen[\rho(k),
\temp]=\beta\fid-\rho\ln(1-\tilde B)-\frac{\tilde D}{\tilde
  B}\ln(1+\tilde B)
\eeq 
where the ideal part is just
\beq
\fid=\kboltz\temp\sum_k \rho(k)\left[\ln\rho(k)-1\right]\label{eq:fid}
\eeq
and we have defined two new parameters $\tilde B$ and $\tilde D$, by
rescaling $B$ and $D$ with the volume
\begin{eqnarray}
\tilde B&=&\frac{B}{\Vol}=\sum_k b_k\rho(k)\\
\tilde D&=&\frac{\beta D}{\Vol^2}=\sum_{k,j}\beta a_{k,j}(\temp)\rho(k)\rho(j)
\end{eqnarray}
The non-ideal part of the free energy in Eq.~\eqref{eq:free_srk} is
called excess free energy $\fexc$ and contains the terms due to the
interaction between the particles in a non-ideal gas.

The Gibbs free energy can now be obtained from the expressions above, simply
by Legendre transforming $\Fen$ as $\Gibbs(\Mol,\pres,\temp)=\min_\Vol\{\Fen(\Mol,\Vol,\temp) +
\pres\Vol\}$. We now introduce a Gibbs free energy per particle
$\gibbs=\Gibbs/\Mol$ and divide again the resulting function into the
ideal part
\begin{equation}\label{eq:gid_srk}
    \beta\gid[\molfrac(k),\pres,\temp]=\frac{\beta G_{\rm id}}{N}=\sum_k\molfrac(k) \ln\molfrac(k) + \ln\beta\pres
\end{equation}
and the excess part
\begin{equation}\label{eq:gexc_srk}
   \beta\tilde{g}[\molfrac(k),\pres,\temp]= \frac{\beta\tilde{G}}{N}=-\sum_k\molfrac(k) \left [\ln\beta\pres\Vol + 1 \right ]+ \frac{\beta\tilde{f}}{\rho}
    +\frac{\beta P}{\rho}
\end{equation}
where the number fraction $\molfrac(k)$ of the specie $k$ is just
$\molfrac(k)=\Mol_k/\Mol=\rho(k)/\rho$, with
$\rho=\sum_k\rho(k)=\Mol/\Vol$ the overall density.

From the above equations~\eqq{eq:gid_srk}{eq:gexc_srk} we can now
derive the phase equilibrium equations $\mu^{a}_k=\mu^{b}_k$
for the coexisting phases $a$ and $b$, and each specie $k$ as
$\mu_k=\partial\Gibbs/\partial\Mol_k=\partial\gibbs/\partial\molfrac(k)$.
For a system of $M$ species, dividing in $P$ phases, the phase
equilibrium is therefore fully analyzed by solving a system of
$(P-1)M$ equations, plus the $M$ equations given by the conservation
of the total number of particles, \ie,
$\sum_a\molfrac^a(k)=\molfrac^{(0)}(k)$, where $\molfrac^{(0)}(k)$ is
the number density of the $k$th specie of the parent, in the $MP$
unknowns $\molfrac^a(k)$. The values of $\pres$ and $\temp$ are set 
as external parameters.

As far as the Peng-Robinson equation of state is concerned, not much
change is needed in the equations above. The PR equation is generally
written as 
\begin{equation}\label{eq:pr}
    P=\frac{\Mol\kboltz\temp}{V-B}-\frac{D}{V(V+B)+B(V-B)}
\end{equation}
where $B$ and $D$ differ from the SRK case in the numerical
coefficients of $b_k$ and $a_{C,k}$. Once again, from the equation
above one gets the excess part of Helmoltz free energy as
\beq\label{eq:fexc_pr}
\beta\fexc=\rho\ln\frac{\Vol}{\Vol-B}+\frac{\sqrt
  2}{4}\frac{D}{B\Vol}\ln\left (
  \frac{V+(1-\sqrt{2})B}{V+(1+\sqrt{2})B} \right )
\eeq
Similarly, the excess part of the Gibbs free energy per particle turns
out to be
\beq\label{eq:gexc_pr}
\beta\tilde{g} = -\sum_k \molfrac(k)\left [  \ln\beta P V +1 \right ] + \frac{\beta\tilde{f}}{\rho}
    +\frac{\beta P}{\rho} 
\eeq
\subsection{Truncatable systems}
A polydisperse system is said to be {\em truncatable} when the excess
part of its free
energy, say the Helmoltz free energy, is a function a limited number
of moments $\rho_i$ of the density distribution $\rho(k)$
\beq\label{eq:moms}
\rho_i=\sum_k w_i(k)\rho(k)
\eeq
with given weight functions $w_i(k)$. In other words, the Helmoltz
free energy of a truncatable system is 
\[
\fen[\rho(k),\temp]=\sum_k\rho(k)\left[\ln\rho(k)-1\right]+\fexc(\rho_i)
\]
For a truncatable system one has
\[
\beta\mu(k)=\frac{\partial(\beta\fen)}{\partial\rho(k)}=\ln\rho(k)+\sum_i
w_i(k)\beta\tilde\mu_i
\]
where the excess moment chemical potentials
$\tilde\mu_i=\partial\fexc/\rho_i$. By imposing the equality of the
chemical potentials in all the coexisting phases, one gets that the
density distribution of the coexisting phases must have the form
\beq\label{eq:dens_sol}
\rho^a(k)=R(k)\exp[-\beta\sum_i\tilde\mu_i^a w_i(k)]
\eeq
By imposing the lever rule, \ie, the conservation of the total number
of particles per specie, $\sum_a\vol^a\rho^a(k)=\rho\parent(k)$
($\rho\parent(k)$ is the density of the parent and
$\vol^a=\Vol^a/\Vol$ is the volume occupied by the phase) one
finds that the function $R(k)$ has the form 
\beq\label{eq:parent_real}
R(k)=\frac{\rho\parent(k)}{\sum_a\vol^a\exp[-\beta\tilde\mu^a(k)]}=\frac{\rho\parent(k)}{\sum_a\vol^a\exp[-\beta\sum_i\tilde\mu^a_i
  w_i(k)]}
\eeq
Although formally solved through the two equations above, an actual
numerical solution of the system is not easily
found. Eq.~\eqref{eq:dens_sol} actually represents a set of, say, $M$
(for $M$ species of particles) self consistent
all strongly coupled through the denominator of Eq.~\eqref{eq:parent_real}.
The problem is that, although the excess free energy is a
function just of the moments $\rho_i$, usually a much smaller
number than the number of species, the ideal part of the free energy is still function
of the whole density distribution $\rho(k)$. Ideally, one would like to reduce the
problem to a smaller number of degrees of freedom, by expressing also the
ideal part of the free energy as a function of the moments only. While
this argument will be treated in the next section, here we will show
that both the SRK and the PR equations of state are in fact truncatable.

In fact, it is rather easy to show that the equations of state
generate two truncatable systems if one introduces two moments of the
density distribution $\rho_1$ and $\rho_2$ as follows
\begin{eqnarray}
\rho_1&=&\tilde B=\sum_k b_k\rho(k)\label{eq:rho1}\\
\rho_2&=&\sum_k d_k\rho(k)\label{eq:rho2}
\end{eqnarray}
where, from Eq.~\eqref{eq:a_poly}, $d_k=\sqrt{\beta
  a_{C,k}\alpha_k(\temp)}$. From the definition above and
Eq.~\eqref{eq:D_poly}, we get that $\tilde
D=\rho_2^2$. Plugging~\eqq{eq:rho1}{eq:rho2} into
Eq.~\eqref{eq:free_srk}, we therefore get, for the SRK equation of states
\beq\label{eq:free_srk_mom}
\beta\fexc(\rho, \rho_1, \rho_2)=-\rho\ln(1-\rho_1)-\frac{\rho_2^2}{\rho_1}\ln(1+\rho_1)
\eeq
Note that in the above expression, the overall density $\rho$ is itself a moment
of the density distribution, with weight function $w_0(k)=1$, as
$\rho=\rho_0=\sum_k\rho(k)$. In other words, the excess part of the
free energy is fully described by the knowledge of only three moments
of the density distribution $\rho_0,\rho_1,\rho_2$ and not on the
whole distribution $\rho(k)$ itself.

For the PR equation of state the argument is again similar to the
case of the SRK equation of states. Introducing again the three
moments $\rho_0,\ \rho_1$ and $\rho_2$, we get that the excess part of
the Helmoltz free energy is just
\beq\label{eq:free_pr_mom}
\beta\fexc(\rho_0, \rho_1, \rho_2)=-\rho\ln(1-\rho_1)-\frac{\sqrt{2}}{4}\frac{\rho_2^2}{\rho_1}
    \ln \left ( \frac{1+(1-\sqrt{2})\rho_1}{1+(1+\sqrt{2})\rho_1} \right )
\eeq
As far as the Gibbs free energy is concerned, it is easy to
show~\cite{SolWarCat01} that the Gibbs free energy inherits its moment
structure from the Helmoltz free energy. However, this time, one
usually introduces normalized moments $m_i$ of the number density
distribution $\molfrac(k)$ defined as
\beq\label{eq:norm_mom}
m_i=\sum_k w_i(k)\molfrac(k)=\frac{\rho_i}{\rho_0}
\eeq
Clearly this time $m_0=\sum_k\molfrac(k)=1$, thus, since $\fexc$
depends on three moments, $\rho_0,\ \rho_1,\ \rho_2$, the Gibbs free
energy depends itself on the overall density, which, however, is
obtained from the equation of state, as a function of $\pres,\ m_1$
and $m_2$. In other words, the Gibbs
free energy turns out to have one degree of freedom less than the
Helmoltz free energy.

In any case, with the definitions above, one gets that the excess part
of the Gibbs free energy for the SRK equation of states is just
\beq\label{eq:gibbs_srk_mom}
\beta \tilde{g}(m_1,m_2)=\ln \frac{\rho_0}{\beta P}-1 +\beta\frac{P}{\rho_0}- \ln(1-\rho_0
m_1)-\frac{m_2^2}{m_1}\ln(1+\rho_0 m_1)
\eeq
Similarly, for the PR equation of state, we get
\beq\label{eq:gibbs_pr_mom}
  \beta \tilde{g}(m_1,m_2) = \ln \frac{\rho_0}{\beta P}-1 +\beta\frac{P}{\rho_0} - \ln(1-\rho_0
    m_1)+\frac{\sqrt{2}}{4}\frac{m_2^2}{m_1}
    \ln \left ( \frac{1+(1-\sqrt{2})\rho_0 m_1}{1+(1+\sqrt{2})\rho_0 m_2} \right )
\eeq
Note that in the two equations above we have omitted the dependence of
$\gexc$ on $\pres$ and $\temp$.

\section{The moment method}
Truncatable systems allow to express the excess free energy as a
function of a small, say $M$, number of moments only. However, as we saw in the
previous section, the difficulty of solving the phase coexistence
equations remains largely unaltered, as the ideal part of the free
energy is still function of the full density (or number)
distribution. Ideally, one would like to express the ideal free energy
too, as a function of the moments only. This is in fact possible, by
means of the moment
method~\cite{ClaCueSeaSolSpe00,SolCat98,SolWarCat01}. While the
following description will refer mostly to the Helmoltz free energy,
similar considerations can be made for the Gibbs free
energy~\cite{SolWarCat01}, with the introduction of the normalized
moments and the number density distribution.

The moment method arises from the hypothesis, in fact verified in
different
works~\cite{ClaCueSeaSolSpe00,SolCat98,SpeSol_p2,SpeSol_p2_fat}, that
the excess free energy is mostly responsible for the phase behaviour
of the whole system. This is in fact not surprising, as the ideal 
free energy is overall convex, and thus does not allow for phase
separation. With this in mind, the moment free energy is constructed
as follows. We subtract from the actual free energy a term
$\rho(k)\ln\rho\parent(k)$, where $\rho\parent(k)$ is the density
distribution of the parent. This term, as linear in the density
$\rho(k)$, does not affect the phase behaviour, as it adds just a
constant to the chemical potential
$\mu(k)=\partial\fen/\partial\rho(k)$. The resulting function is then
minimized with respect to $\rho(k)$ with the $M$ moments appearing in
the excess part as constraints (2 in the two previous cases). The
minimum value of the resulting free energy is then found to be
\beq\label{eq:fmom_gen}
\fmom(\vectrho)=\sum_i^M\lambda_i\rho_i-\rho_0+\fexc(\vectrho)
\eeq
where $\vectrho$ is just a vector having the moments $\rho_i$ as
components and the $\lambda$s are the $M$ Lagrange multipliers. 
The minimum value of the free energy is reached for a density
distribution from the family
\beq\label{eq:dens_mom_gen}
\rho_{\rm mom}(k)=\rho\parent(k)\exp\left(\sum_i\lambda_iw_i(k)\right)
\eeq
From the moment free energy~\eqref{eq:fmom_gen}, one can also define
the moment chemical potentials $\mu_i$, as
$\mu_i=\partial\fmom/\partial\rho_i=\lambda_i+\tilde\mu_i$. The
pressure is obtained from the Gibbs-Duhem relation as 
\[
\pres=\sum_k\mu(k)\rho(k)-\fen=\sum_i\mu_i\rho_i-\fen
\]
It is easy to show that the above expression obtained from the
moment free energy is in fact identical to the one obtained from the
exact free energy~\cite{SolWarCat01}. Furthermore, it is easy to show
that the moment free energy correctly detects the 
onset of phase coexistence. In other words, one can show~\cite{SolWarCat01,Sper02} that any two
phases coexist for the full system, if and only if, they coexist for the
moment free energy, thus, $\mu\ph(k)=\mu^b(k)\Leftrightarrow\mu_i\ph=\mu^b_i$.
Thus, at least up to the onset of the phase coexistence,
the full solution in Eq.~\eqref{eq:dens_sol} actually belongs to
the family in Eq.~\eqref{eq:dens_mom_gen}. Cloud point
and shadow phases are then correctly detected by the moment free energy
solution above. Furthermore, spinodals and critical/tricricital points
are found exactly~\cite{SolWarCat01}.


The enforcement of the lever rule only for the moments is in fact the only
approximation we make in using the moment method, as this does not
ensure the satisfaction of the complete levere rule, while, as
mentioned earlier, equality of pressure and chemical potentials are
ensured. However, as shown in details
in~\cite{ClaCueSeaSolSpe00,SpeSol_p2,SpeSol_p2_fat}, the approximation
can be reduced efficiently by retaining extra moments and, in
particular, by means of the adaptive method of choice of extra weight
functions which allows to obtain a solution as close as wanted to the
exact one, by retaining only two extra moments.

In order to give a more precise insight of the actual problem one has to
solve in the case of the two equations of state mentioned earlier, let
us sketch the resulting system of equations, obtained within the
moment method approach. As mentioned, since we have $\pres$ and
$\temp$ as external parameters, we move on to the Gibbs
formalism. Thus we evaluate the gibbs free energy and from that, we
calculate the moment chemical potentials as $\mu_i=\partial\gibbs_{\rm
  mom}/\partial m_i$, where $m_i$ are the normalized moments. Let us
now assume we have a Gas-Liquid demixing and let us call
$\phi\ph=\Mol\ph/\Mol\parent$ the fraction of particles in each
phase (G or L). The phase coexistence is then fully solved by
enforcing the equality of the moment chemical potentials and of the
quantity $\Pi=\gibbs_{\rm mom}-\ln\pres-\sum_{i\neq 0} m_i\mu_i$,
which is a sort of Legendre transform of the
pressure~\cite{SolWarCat01}, in all the coexisting phases. We must also enforce the
conservation of the total number of particles, \ie,
$\sum_a\Mol\ph(k)=\Mol\parent(k)$. If we multiply by $w_i(k)$ on both
sides and sum over $k$, we get, after rearranging, the lever rule for
the normalized moments $\sum_a\phi\ph m\ph_i=m\parent_i$, which is
the condition we actually enforce. Thus, for the two equations of
state, the system of equations we have to solve turns out to be
\[
\left\{
\begin{array}{lcl}
\mu_1\G&=&\mu_1\L\\
\mu_2\G&=&\mu_2\L\\
\Pi\G&=&\Pi\L\\
m\parent_1&=&\phi\G m_1\G + (1-\phi\G) m_1\L\\
m\parent_2&=&\phi\G m_2\G + (1-\phi\G) m_2\L\\
\pres&=&\pres(\rho_0\G,m_1\G,m_2\G)\\
\pres&=&\pres(\rho_0\L,m_1\L,m_2\L)
\end{array}
\right.
\]
\ie, 7 equations in the 7 unknowns
$\lambda_1\G,\lambda_1\L,\lambda_2\G,\lambda_2\L,\rho_0\G,\rho_0\L, \phi\G$.
The moment chemical potentials and the pressure are then for
the SRK equation of state:
  \begin{eqnarray*}
    \beta \mu_1 &=& \beta \lambda_1 +\frac{\rho_0}{1-\rho_0 m_1} -
    \frac{\rho_0 m_2^2}{m_1(1+\rho_0 m_1)}+\frac{m_2^2}{m_1^2}
  \:\ln\left(1+\rho_0 m_1\right)\\
  \beta \mu_2 &=& \beta \lambda_2 -2\:\frac{m_2}{m_1}\:\ln\left(1+\rho_0
    m_1\right)\\
  \beta\pres&=&\frac{\rho_0}{1-\rho_0 m_1}-\frac{\rho_0^2 m_2^2}{1+\rho_0 m_1} 
  \end{eqnarray*}
For the PR equation of state, the calculations are just slightly more
complicated, as now we get:
\begin{eqnarray*}
  \beta \mu_1 &=& \beta \lambda_1 +\frac{\rho_0}{1-\rho_0 m_1} -
  \frac{\rho_0 m_2^2}{m_1(1+2\rho_0 m_1-\rho_0^2 m_2^2)}-\frac{\sqrt{2}}{4}\:\frac{m_2^2}{m_1^2}
  \:\ln\left[\frac{1+(1-\sqrt{2})\rho_0 m_1}{1+(1+\sqrt{2})\rho_0 m_1}\right]\\
  \beta \mu_2 &=& \beta \lambda_2 +\frac{\sqrt{2}}{2}\:\frac{m_2}{m_1}
  \:\ln\left[\frac{1+(1-\sqrt{2})\rho_0 m_1}{1+(1+\sqrt{2})\rho_0
      m_1}\right]\\
  \beta\pres&=&\frac{\rho_0}{1-\rho_0 m_1}-\frac{\rho_0^2
    m_2^2}{1+2\rho_0 m_1-\rho_0^2 m_1^2}
\end{eqnarray*}

\section{Numerical results}

As a way of example, in order to show the potential of the method
described in the previous section, in this section we will show some
numerical results obtained by applying it to a real case-study. We
solve the phase equilibria for a mixture of 24 hydrocarbons, up to
nC15, along a straight line crossing the $(P,T)$ phase diagram. The
calculations are done using the SRK equation of state, although no
relevant difference is observed when using the PR equation of state. 
Our numerical results are compared with the results obtained with a commercial
program (PVTsim), licensed to Snamprogetti s.p.a. (that provided the
results). In Fig.~\ref{fig:C1_kashagan}, we plot the concentration in
mole \% of C1 (methane), in the two coexisting phases, against the
temperature. 
\bfig
\label{fig:C1_kashagan}
\begin{picture}(0,0)%
\includegraphics{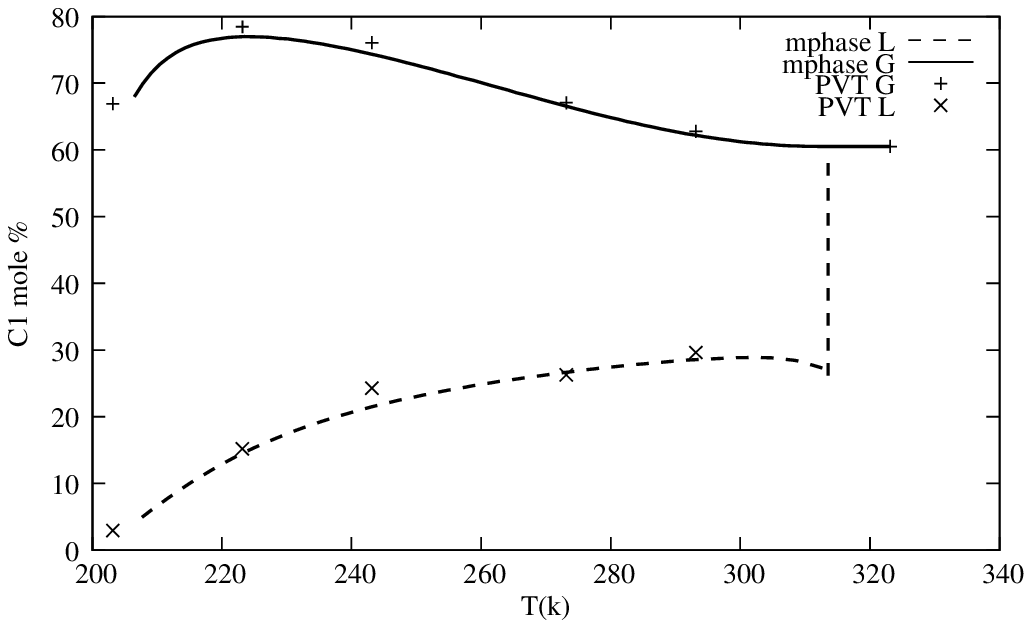}%
\end{picture}%
\setlength{\unitlength}{3947sp}%
\begingroup\makeatletter\ifx\SetFigFont\undefined%
\gdef\SetFigFont#1#2#3#4#5{%
  \reset@font\fontsize{#1}{#2pt}%
  \fontfamily{#3}\fontseries{#4}\fontshape{#5}%
  \selectfont}%
\fi\endgroup%
\begin{picture}(4947,2977)(1226,-3424)
\end{picture}%
\caption{Concentration of methane against temperature in the two
  coexisting phasese (solid and dashed lines), compared to the results
  obtained with PVTsim (diamonds). The liquid phase appears correctly
  at lower temperature as the path across the phase diagram crosses
  the phase envelope. The cloud point is detected exactly by our
  method. Deep inside the coexistence region some small deviations
  appear, although it is not clear whether they are due to the moment
  method, or to the approximations introduced by PVTsim.}
\vspace*{-0.5cm}
\efig
Our result agrees very well with the points obtained with PVTsim. The
concentration of both phases is evaluated correctly. Furthermore, the
cloud point, \ie, the point at which the liquid phase appears, is
detected exactly on the phase envelope shown by PVTsim.

In Fig.~\ref{fig:C4_kashagan} we show again the same case. Now we plot
the concentration of n-C4 against the pressure. As the
pressure-temperature path enters the coexistence region, liquid is
found. This time the component represents less than 1\% of the total
composition of the gas, while it is about 10\% of the total
composition of the liquid. Again, even with a heavier hydrocarbon, our
numerical results are in excellent agreement with the ones obtained
with PVTsim.
\bfig
\label{fig:C4_kashagan}
\begin{picture}(0,0)%
\includegraphics{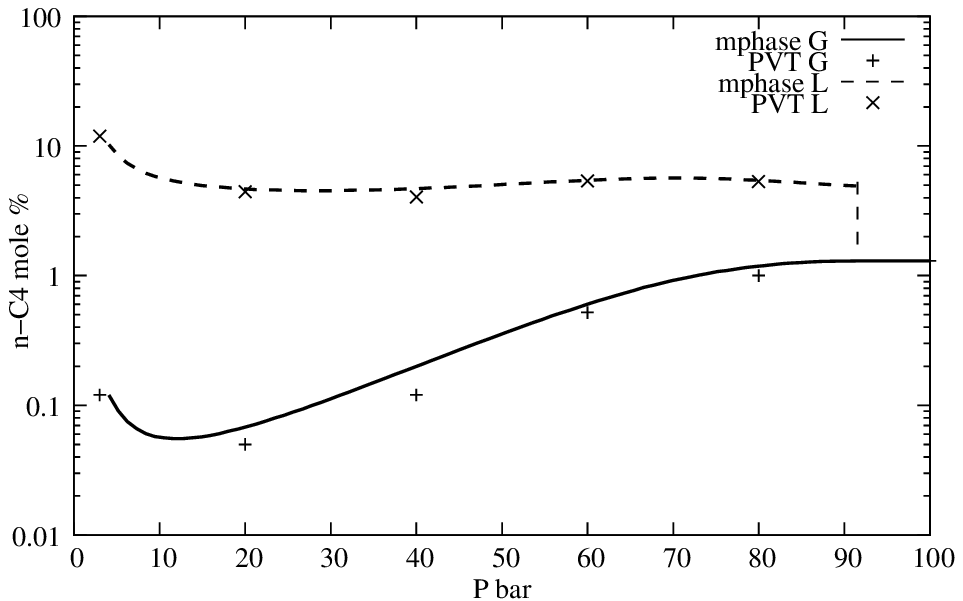}%
\end{picture}%
\setlength{\unitlength}{3947sp}%
\begingroup\makeatletter\ifx\SetFigFont\undefined%
\gdef\SetFigFont#1#2#3#4#5{%
  \reset@font\fontsize{#1}{#2pt}%
  \fontfamily{#3}\fontseries{#4}\fontshape{#5}%
  \selectfont}%
\fi\endgroup%
\begin{picture}(4638,2885)(1405,-3300)
\end{picture}%
\caption{Concentration in mole \% of n-C4 against pressure across the
  phase diagram. As pressure and temperature drop enough to enter the
  phase coexistence region, the liquid phase is found. The
  concentration the element in both phases are in excellent agreement
  with the ones obtained with PVTsim. Again, it is not clear whether
  the small deviations inside the coexistence region are due to our
  method, or rather to approximations and truncations introduced by
  PVTsim.}
\vspace*{-0.8cm}
\efig
\vspace*{-0.3cm}
\section{Conclusion}
We have applied the moment method to the analysis of phase equilibria
of mixture of hydrocarbons, using the SRK and PR equations of state. 
Our results show that not only the moment method is
applicable as it correctly detects gas-liquid phase coexistence. Even
with a large number of components in the mixture, our algorithm remains
robust and numerical calculation fast. Furthermore, our numerical
results agree quantitatively very well with the results obtained using a
widely used commercial program (PVTsim). No further demixing,
beyond the G-L coexistence is observed, however, it is not yet clear
whether this depends on the equations of state used, or rather to the
choice of the mixture of hydrocarbons. It may be possible to have the
coexistence of more than two (G and L) phases, \eg, more than one
different liquid and/or gas phases, using a wider distribution of
hydrocarbons. 

As far as future developments are concerned, one could proceed to the
analysis of a fully polydisperse case, \ie, by introducing a
continuous distribution of species. Clearly the two equations should
first be extended to the continuous case, \eg, by introducing a
continuous dependence of the acentric factor on the size of
particles. It is possible that further demixing appears using
different distributions. The extension
to the continuous case should not have any impact on our
method. Clearly, the computation may be slightly slower, as in that case the moments have to
be evaluated by integration rather than summation over a finite number
of species; however, no formal or substantial adjustment is needed.

Finally, further applications of the moment method could be tried. For
instance, one could try to apply it to metallurgy or other cases of
industrial interest. Clearly, when solid phases appear, one would have
to extend the analysis introducing spatial and/or orientational degrees
of freedom. However, as previous results already show~\cite{ClaCueSeaSolSpe00,FasSol03,FasSol04,FasSolSpe04,SpeSol_p2,SpeSol_p2_fat} this should not
represent a limitation for the method. 

\bibliographystyle{plain}
\bibliography{bibliografia}
\end{document}